\documentclass[runningheads,a4paper]{llncs}
\usepackage{amssymb}
\usepackage{graphicx}
\usepackage{epstopdf}
\usepackage{amsmath}
\usepackage{url}
\usepackage{algorithm,algorithmic}
\usepackage{color}
\usepackage{subfig}
\begin{document}

\title{Data Leak Aware Crowdsourcing in Social Network}
%
\author{Iheb Ben Amor \inst{1}, Salima Benbernou \inst{1}, Mourad Ouziri\inst{1}, Mohamed Nadif \inst{1},\\ Athman Bouguettaya\inst{2}}
\institute{Universit\'e Paris Sorbone Cit\'e, Paris Descartes, France 
\and RMIT University, Australia \\ 
\email{firstname.lastname@parisdescartes.fr\\iheb.ben-amor@etu.parisdescartes.fr\\
athman.bouguettaya@rmit.edu.au}}

\maketitle

\begin{abstract}
Harnessing human computation for solving complex problems call spawns the issue of finding the unknown competitive group of solvers. In this paper, we propose an approach called \textit{Friendlysourcing} to build up teams from social network answering a business call, all the while avoiding partial solution disclosure to competitive groups. The contributions of this paper include (i) a clustering based approach for discovering collaborative and competitive team in social network (ii) a Markov-chain based algorithm for discovering implicit interactions in the social network.

\end{abstract}



\keywords{Social network, outsourcing human-computation,privacy }

\section{Introduction}

A new tend of teamwork has been emerged unconstrained by local geography, available skill set, networking and deep  relationships the \textit{crowdsouring}. It is the action of  outsourcing tasks, traditionally performed  by an employee or contractor, to an undefined group of people through an open call \cite{daren_c._brabham_crowdsourcing_2008}. Crowdsourcing applications should be enable to seek for people crowd  on demand  to perform a wide range of complex and difficult tasks. Thousands human actors will provide their skills and capabilities in response to the call.  
We introduce a type of crowdsourcing  called  \textit{Friendlysouring} based on the efficiency of social network to outsource a task
to be performed by people on demand
instead of  an open world as Mechanical Turk is doing. In fact,  the interactions between people involved to answer a query become complex more and more and the collaboration leads to the emergence of social relations and a social network can be weaved for human-task environment.\\
\textbf{Challenges.}The goal of Friendlysouring system is to let  people collaborating on a joint task in the crowd environment where  they  may seek for other members towards social crowd relationships for achieving a business goal. Thus, many competitive teams can provide a set of answers  to the call. However, as the crowd task is competitive between teams, it is important to group people in a manner  there is no  inter-teams leaking. Such mechanism will avoid  the information  leak  between crowd people in different teams.  Hence,  the issues and challenges considered in our system, include, (i) how to build up and  discover teams answering the query towards the social relationships, (ii) how to avoid the  solution disclosure of the problem during  the teams construction  between competitive groups. 
In fact, people on demand  collaborating to a task  may share sensitive information (part of the problem solution)that may be propagated or forwarded  to other crowd members in the social network.

Few works dealing with crowdousourcing are provided.
In \cite{DBLP:conf/icde/DeutchGKM11}, the  Trivia Masster system  generates a very large Database of facts in a
variety of topics, cleans it towards a game mechanism and uses it for question answering.
In \cite{DBLP:journals/it/SkopikSPTD11} is proposed a novel approach for integrating human capabilities in crowd process flows. In \cite{DBLP:conf/sigmod/2011}, the CrowdDB  system uses human input via crowdsourcing to process queries that neither database systems nor search engines can adequately answer. Privacy and data leaking are not at all discussed in these works. Moreover, privacy have been introduced in social network as in
\cite{DBLP:conf/ccs/FangKLT10} to  design a wizard that may automatically configure a user's privacy settings with minimal effort from the user to aim policy preferences learning.  \cite{DBLP:conf/icde/TalukderOEEY10}, the authors introduced privacy protection tool that measures the amount of sensitive information leakage in a user profile and suggest self-sanitizing action to regulate the amount of leakage. 
The primovoter tools is unable to estimate the leakage based on a private data propagation, so it settle for a direct user connections and an installed applications on friend profiles.

\noindent

\textbf{Contributions.} We address the aforementioned challenges by proposing the Friendlysourcing system to discover data leaking aware competitive teams answering the query through a social network. The discovering method is based on a k-means like algorithm to cluster the potential crowd people from the social network that are \textit{close} to collaborate in the same team. The system will not group them in the clusters that are competitive, thus, avoid inter-teams data leaking. To handle such leak, a Markov chain-based algorithm is proposed to discover the implicit social relationships between crowd people. It knows exactly to whom the user data can be propagated in the social network and hence avoid to let crowd people grouped in different clusters. The approach is based on dynamic model that deals with effective rates of shared data and not only on static friend relationship between crowd members.

The rest of the paper is organized as follows: section 2 provides an overview of our Friendlysourcing system. The propagation process is described in section 3. In section 4 is discussed the clustering based approach to discover competitive clusters in the social network answering a business call. Finally, conclusion and future works are given in the section 5.

\section{Overview of Friendlysourcing Framework}
\begin{figure}[h!]
  \center
  \includegraphics[height=7cm,width=9cm]{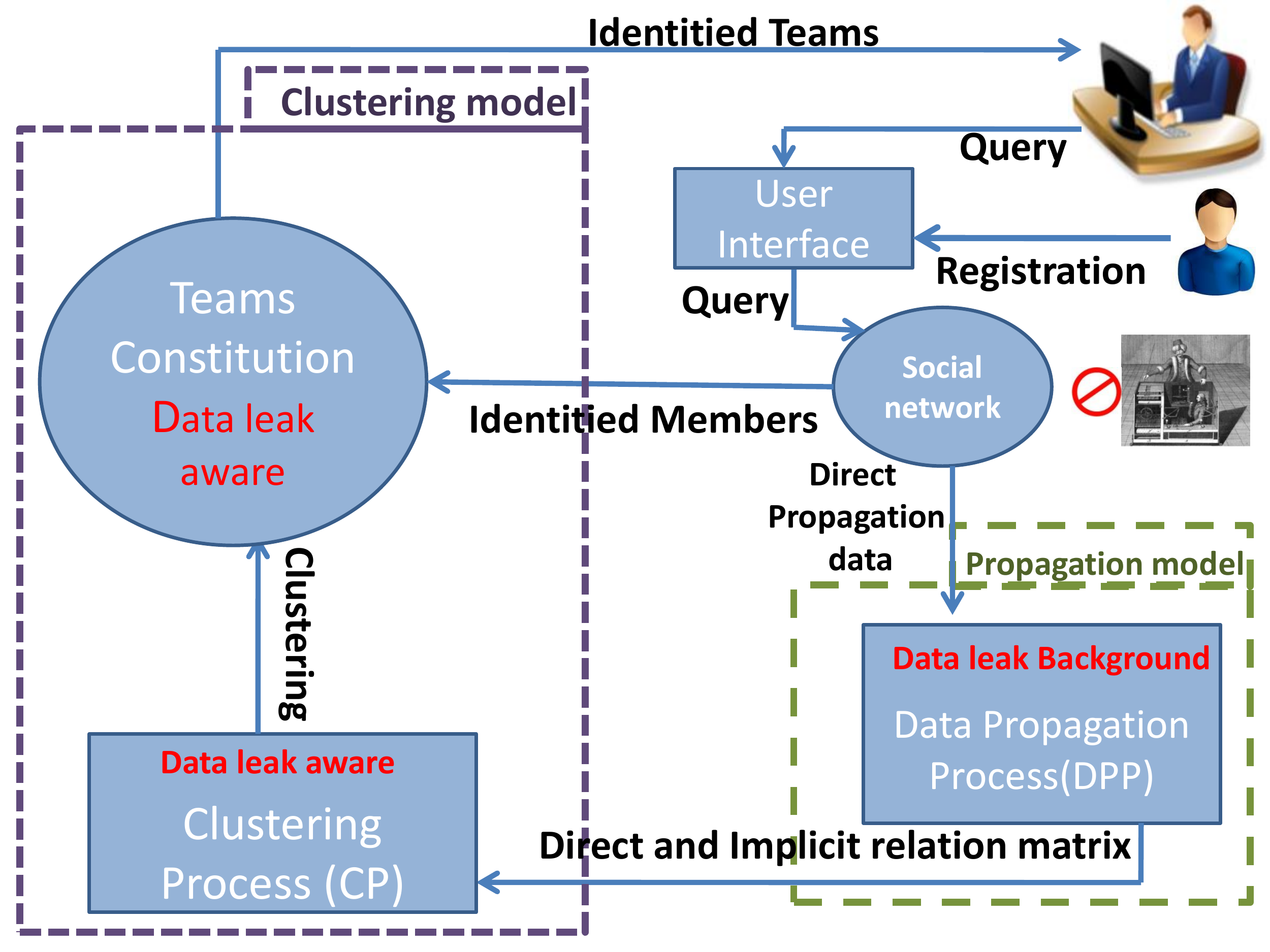}
    \caption{Friendlysourcing architecture}
    \label{archi}
\end{figure}

We devise a crowdsourcing architecture for discovering data leak aware collaborative and competitive teams. It incorporates two main components  discussed in this paper, they are depicted in Figure\ref{archi} and are namely \textit{Data propagation process, clustering process}.
Beforehand, the person responsible of company call may register to the friendlysourcing platform using the user interface. He describes the company activities and submits a query.
Once the call is launched, it will be  visible in our platform.  Every social network member is authorized to register for a call using the user interface. The registered  member can  examine the details of the call and make comments.

\begin{itemize}
\item \textit{Data Propagation Process :}
When the registration is closed, the friendlysourcing system computes the data based on the information collected from the different social networks in order to data leak aware while discovering the teams. In the first step, the request is achieved  by the propagation process, thus identifying the direct relationships between the social network members. After that, the process will discover implicit interactions in the social network and the maximum of data propagation between members. The approach is based on  Markov chain model. The details are provided in the next section.

\item \textit{ Clustering Process}

In a second phase, starting from the whole data  propagation calculated in the propagation process DPP, the Clustering process CP will group the crowd users in the same clusters having strong propagation.
That means, more the relationship probability is higher more the users need to be in the same cluster for the collaboration and not in competitive clusters.

\item \textit{Team Constitution}: The module will  constitute the different team based on the provided result from the clustering process CP. It will use the user profile information provided from the social network. It will notify the users about the team discovering results.

\end{itemize}

\section{A Markov chain-based approach for data propagation}

The social networks is a set of direct relationships between members. These direct relationships allow to compute the probability of data propagation between
only direct friends. However, discovering competitive teams aware of data leak, needs to know all possible interactions. For handling the implicit/indirect relations between members we propose a Markov chain-based approach.


We present in this section a model and an algorithm of data propagation across the entire social network. This allows to compute all indirect interactions between all members and to know to whom the user data can be propagated to.





\subsection{A graph-based model of data sharing relationships}
\label{graph_model}
Our model of the social network is a labeled directed graph $ G \left\langle M, A, P \right\rangle $ where,
\begin{itemize}
   \item $M = \left\{m_i\right\}$: set of nodes where each node represents a member of social network.
   \item $A = \left\{a_{ij}=(m_i,m_j) / (m_i,m_j) \in M \right\}$: set of edges where each edge represents a direct friend relationship between two members.
   \item $P = \left\{p_{ij} /\forall i,j ~ p_{ij} \in \left[0,1\right] \right\}$: is set of labels where each label $p_{ij}$ of the  edge $a_{ij}$ represents the rate/probability of data shared by the member $m_i \in M$ with his friend member $m_j \in M$.
\end{itemize}
In the given graph model, \textit{friend} relationship is represented using edge $A$ labeled with the real probability of shared data $P$.

The probability $p_{ij}$ that the member $m_i$ shares owned data with member $m_j$ is computed in real time using the following formula:

$         p_{ij} =  \frac{quantity ~ of ~ data ~ that ~ m_i ~ shared ~ with ~ m_j}{quantity ~ of ~ data ~ held ~ by ~ m_i}$  \\

Let's consider the example of a social network depicted in Figure \ref{fig_graph_exple}:
\begin{figure}
    \center
    \includegraphics[height=5cm,width=8cm]{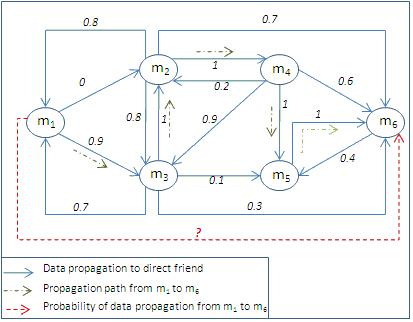}
    \caption{Example of interactions between crowd members in the social networks}
    \label{fig_graph_exple}
\end{figure}
\begin{itemize}
   \item the arc $(m_1,m_3)$ indicates that $m_1$ has friend relationship with $m_3$, and shares with him 90\% of his data.
   \item the arc $(m_1,m_2)$ indicates that $m_1$ has friend relationship with $m_2$ but he never shares with him any data.
\end{itemize}

The presented graph-based model is a set of direct relationships between members. These direct relationships provide the probability of data propagation between only direct friends. We present in the following the Markov propagation model to compute the probability of data propagation between indirect friends (such as propagation rate from $m_1$ to $m_6$ in Figure \ref{fig_graph_exple} ).

\subsection{Markov chain-based model for data propagation}
\label{MarkovModel}

Given an owned data of a member, we propose a Markov chain-based model to compute the propagation probability of this data in the entire social newtork.

In social networks, data is propagated from friend to friend following a Markov chain model \cite{MarkovBasics}. That is, a social network member shares owned data only with his friends and, then, each friend shares the data with only their friends and so on.

\begin{definition}
A Markov chain is a sequence of random variables $ X_1,...,X_n $ with the Markov property, namely that, the future state depends only on the the present state, and not on the past states.
Formally,

$ P(X_{n+1}=x|X_1=x_1, X_2=x_2, \ldots, X_n=x_n) = P(X_{n+1}=x|X_n=x_n) $

\end{definition}

From this formal definition, the probability that a given member get a data depends on probability to get it from only his direct friends (and not from indirect friends).

The probability of data propagation between direct friends may be represented with Propagation Matrix defined as follows:

\begin{definition}
\label{def_prog_matrix}

Propagation Matrix of a social network is matrix that gives probability $ p_{ij}$ of propagating data between each couple of members $ (m_i,m_j) $:

\[ \bordermatrix{
       ~   & m_1    & m_2    & m_3    & ...  & m_n \cr
       m_1 & p_{11} & p_{12} & p_{13} & ...  & p_{1n}\cr
       m_2 & p_{21} & p_{22} & p_{23} & ...  & p_{2n}\cr
       m_3 & p_{31} & p_{32} & p_{33} & ...  & p_{3n}\cr
       ... & ...    & ...    & ...    & ...  &  ... \cr
       m_n & p_{n1} & p_{n2} & p_{n3} & ...  & p_{nn}
}
\]

where

$$
p_{ij} = \left\{
    \begin{array}{ll}
       \frac {quantity ~ of ~ data ~ that ~ m_i ~ shared ~ with ~ m_j}{quantity ~ of ~ data ~ held ~ by ~ m_i} & \mbox{if } (m_i,m_j) \in A \\
       1 & \mbox{for } i=j \\
       0 & \mbox{else}
    \end{array}
\right.
$$

\end{definition}

This propagation matrix has the following properties:
\begin{itemize}
   \item $ p_{ii} = 1$, which means that member $m_i$ does not lost owned data when he shares it.
   \item $ \sum_{k\in\left[1,n\right]}\left( p_{ik} \right) \neq 1 $, because data may be propagated to several members at the same time.
   \item $ p_{ij} \neq p_{ji} $, which means that a member $m_i$ may share with a friend $m_j$ a quantity of data different his friend $m_j$ may share with him.
   \item $ \exists (i, j) | (m_i,m_j) \in A \wedge p_{ij} = 0 $, which means that members do not share necessarily data with their friends.
   \item if $m_i$ and $m_j$ are not direct friends then $p_{ij}$ = 0 .
\end{itemize}

The propagation matrix of the social network of figure \ref{fig_graph_exple} is given as follows:
\[ \bordermatrix{
       ~   & m_1 & m_2 & m_3 & m_4 & m_5 & m_6  \cr
       m_1 & 1   & 0   & 0.9 & 0   & 0   & 0    \cr
       m_2 & 0.8 & 1   & 0.8 & 1   & 0   & 0.7  \cr
       m_3 & 0.7 & 1   & 1   & 0   & 0.1 & 0.3  \cr   
       m_4 & 0   & 0.2 & 0.9 & 1   & 1   & 0.6  \cr  
       m_5 & 0   & 0   & 0   & 0   & 1   & 1    \cr  
       m_6 & 0   & 0   & 0   & 0   & 0.4 & 1    \cr          
}
\]

\subsection{A Markov chain-based algorithm of data propagation}

The propagation matrix of section \ref{graph_model} gives only probability of data propagation between direct friends.

But it is not sufficient to compute the probability that data is propagated from member to indirect-friend because:
\begin{enumerate}
   \item Propagation matrix defined in Definition \ref{def_prog_matrix} does not give the real propagation probabilities between members.
In Figure \ref{fig_graph_exple}, the direct propagation probability from $m_1$ to $m_2$ is zero ($p_{12} = 0$). However, through $m_3$, data of $m_1$ may be propagated to $m_2$ with probability $0.9 \times 1$ $= 0.9$.
   \item The Propagation matrix does not provide the data propagation to indirect-friends. It's indicate a zero value of sharing data with indirect friends,
    because members share their data only with direct friends. In the propagation matrix of figure \ref{fig_graph_exple}, probability that data of member $m_1$ may be propagated to his indirect-friends $m_5$ and $m_6$ is zero because $m_1$ is no direct friend relationship with them. However, data may be propagated from  $m_1$ to $m_5$ through $m_3$ with probability $0.9 \times 0.1 = 0.09 $.
   \item Propagation to indirect-friends is hard to calculate: as example, what is the probability that data of $m_1$ may be propagated to $m_6$ (probability of dotted red arrow in figure \ref{fig_graph_exple})? To calculate this probability, we have to explore all the paths allowing propagation of data from $m_1$ to $m_6$. Each one allows to calculate a propagation probability. The final propagation probability is the maximum of propagation probability of all the possible paths, wihich corresponds to the propagation risk. The path $(m_1, m_3, m_2, m_4, m_5, m_6)$ indicated with dotted green arrows in figure \ref{fig_graph_exple} allows propagation of data from $m_1$ to $m_6$ with the maximum probability $0.9 \times 1 \times 1 \times 1 \times 1 = 0.9$.
   However, it is hard to calculate this probability because real social networks are complex.
\end{enumerate}

For this reasons, we need to design an efficient algorithm that calculates the optimal data propagation probability from the owner to all the members of the social network. This algorithm is based on energy function we define as follows:

\begin{definition}
The energy function $ p_{i} $ of member $ m_i $ is the probability that data is propagated to member $ m_i $. In our model, data is propagated following Markov chain. That is:

\begin{center}
$ p_i = \underset{m_k\in N_{m_i}}{Max}(p_k \times p_{ki}) $  ~~~~~~~~~~~~~~~~~~~~~~ $(1)$
\end{center}

\end{definition}
where,
\begin{itemize}
   \item $ N_{m_i}  $ is a set of direct friends of $m_i$,
   \item $ p_k      $ is the energy function of $m_k$,
   \item $ p_{ki}   $ is the probability of propagating data from $m_k$ to $m_i$.    
\end{itemize}

To compute the energy function $p_i$ for all members $m_i$ of the social network, we have to use an iterative algorithm \cite{W2009GOEB}. We design our simple algorithm Algorithm \ref{algo_propagation}.

\begin{algorithm}[h]
\caption{\textsc{Propagation probability computing algorithm}}
\label{algo_propagation}

\begin{algorithmic}[1]
\REQUIRE $ G \left\langle M, A, PM \right\rangle $ -- labeled directed graph of the social network where $ PM $ in the propagation matrix\\
   \hspace{0.9 cm}    $ m_{ow}$ -- owner of the data\\
   \hspace{0.9 cm}    $ m_r   $ -- recipient member of data that we want calculate the propagation probability \\
   
\ENSURE $ p_r $ -- probability that owned data is propagated to  $m_r$

\PRINT $ \mathcal{P} = (p_1, \ldots, p_{ow}, \ldots, p_r, \ldots , p_n) $: Energy function at the previous step.

\PRINT $ \mathcal{PS} = (ps_1, \ldots, ps_{ow}, \ldots, ps_r, \ldots , ps_n) $: Energy function at the current step.

\PRINT $ continue $: boolean value indicating if the optimum values of all members are reached.

\COMMENT Initializations

\STATE    $ p_{ow} = 1 $ and $ \forall m_i \neq m_{ow}, p_i = 0 $
   
\STATE    $ continue  \leftarrow true $

\COMMENT Iterations
   
\WHILE{$continue$}         
       \FOR {each members $ m_i \neq m_{ow} $}                                
                   \STATE $ ps_i = Max_{m_k\in N_{m_i}}(p_k \times p_{ki})) $                    
       \ENDFOR
           
       \IF {$ \mathcal{P} \neq \mathcal{PS} $}            
                   \STATE$ \mathcal{P} \leftarrow \mathcal{PS} $            
       \ELSE                
                   \STATE $ continue \leftarrow false $                    
       \ENDIF            
\ENDWHILE

\RETURN $ p_r $
\end{algorithmic}
\end{algorithm}

The algorithm processes as follows:
\begin{enumerate}
   \item Initialisations: $p_{ow}=1,$ $\forall i\neq ow$ $ p_i=0 $. That is, only the owner $m_{ow}$ has the data.
   \item Iterations: At each iteration, the algorithm computes $p_i$ for $m_i \in N_{m_i}$ using formula (1).
   \item Stop: The algorithm stops when the probability maximum  of each member is reached.
\end{enumerate}

Applying this algorithm on the propagation matrix of section \ref{MarkovModel}, we get the following matrix completed with indirect-friend propagation probabilities:
\[ \bordermatrix{
       ~   & m_1 & m_2 & m_3 & m_4 & m_5 & m_6  \cr
       m_1 & 1   & 0.9   & 0.9  & 0.9   & 0.9   & 0.9    \cr
       m_2 & 0.8 & 1    & 0.8   & 1      & 1      & 1  \cr
       m_3 & 0.8 & 1    & 1     & 1      & 1       & 1  \cr   
       m_4 & 0.7 & 0.9 & 0.9  & 1      & 1       & 1  \cr  
       m_5 & 0   & 0   & 0       & 0      & 1       & 1    \cr  
       m_6 & 0   & 0   & 0       & 0      & 0.4    & 1    \cr          
}
\]

\section{Data disclosure aware clustering process}
Based on data propagation calculated in the previous section, the clustering process groups of crowd members into free data leak clusters.

\begin{definition}
A cluster $C$ is set of crowd members having or no strong propagation:
\begin{center}
$C = \left\{ m_i \right\}$ such that $\forall m_i,m_j \in C, p_{ij}\in [0,1], p_{ji} \in [0,1]  $
\end{center}
\end{definition}

\begin{definition}
\label{def_leakage}
Two clusters $C_k$ and $C_s$ are free data leak iff:
\begin{center}
$\forall m_i \in C_k, \forall m_j \in C_s, p_{ij} \leq \eta \wedge p_{ji} \leq \eta $
\end{center}
\end{definition}

From the definition \ref{def_leakage}, we consider there is a risk of data leak between two clusters $C_k$ and $C_s$ if the propagation rate between all the members of the two clusters is less than a threshold $\eta$. The later is proposed as a value for which the data propagation of a crowd member is acceptable in a social network. The dynamacity of social member profil impactes the value of $\eta$, but it is out of the scope of the paper.

The propagation matrix calculated by the algorithm \ref{algo_propagation} is be updated as folllows:
\begin{center}
$\forall i,j, p_{ij}=Max(p_{ij}, p_{ji})$
\end{center}

The propagation matrix of our example is updated as follows:
\[ \bordermatrix{
       ~   & m_1 & m_2 & m_3 & m_4 & m_5 & m_6  \cr
       m_1 & 1   & 0.9 & 0.9 & 0.9   & 0.9   & 0.9    \cr
       m_2 & 0.9 & 1   & 1   & 1   & 1   & 1  \cr
       m_3 & 0.9 & 1   & 1   & 1 & 1 & 1  \cr   
       m_4 & 0.9   & 1   & 1 & 1   & 1   & 1  \cr  
       m_5 & 0.9   & 0   & 1 &1   & 1   & 1    \cr  
       m_6 & 0.9   & 0   & 1 & 1 & 1   & 1    \cr          
}
\]

Based on this updated propagation matrix, the members are classified into free leak clusters using a clustering algorithm.

Our clustering algorithm is a k-means algorithm \cite{mcqueen1967}. The principle of the algorithm is that for each cluster $C_k$ and each member $m_i$, the member $m_i$ is calssified to the cluster $C_k$ if there is high data propagation between the member $m_i$ and at least one of the members of $C_k$. The algorithm uses the following specific distance called $D_{max}$:
\begin{center}
$ D_{max}(C_k,m_j) = \underset{m_i\in C_k} {Max}~ p_{ij} $
\end{center}

where $Max$ is the maximum function, $C_k$ a cluster to be built, $m_j$ is member that can be clustered into the cluster $C_k$, and $m_i$ is a crowd member in $C_k$, $p_{ij}$ is the propagation value between $m_i$ and $m_j$ given in te propagation matrix.

The clustering algorithm prossess as follows:

\begin{itemize}
\item Inputs:
the data disclosure threshold $\eta$, number of clusters.
\item Initialization:
The initialization of the clusters is done by assigning arbitrarily a member to each cluster.
\item Iterations:
For each candidate member $m_i$ and a cluster $C_j$, if $ D_{max}(C_j,m_i) \geq \eta$ then $m_i$ is added to $C_k$
\item Stop:
The algorithm is deemed to have converged when the assignments of members to clusters no longer change.
\end{itemize}


Moreover, in the case that a candidate member has a strong communication with more than one cluster, we will merge the clusters with whom the candidate member has a high data propagation and assign it to the new merged cluster. Because it may probably disclosure the data of the cluster to the other clusters. For instance if $d(C_1,m_{k}) = 0,8$ and $d(C_2,m_{k}) = 0,7$, then we will merge the cluster $C_1$ and $C_2$ and K will integrate the cluster $C_{12}$ result of the $C_1$ and $C_2$ fusion. The algorithm is presented as Algorithm 2.
\begin{algorithm}[h]
\caption{\textsc{D-Max discovering teams algorithm}}
\label{algo_k-means}

\begin{algorithmic}[2]
   

\PRINT $ Clusters=(Clust_{1},Clust_{2},...,Clust_{Cluster})$: Teams constitution

\COMMENT Initializations

\STATE    $ Distances,Centroid,MaxDistancesMax = 0 $
\STATE    $ ClustNB = -1 $
\STATE  $ Threshold = \eta $
\FOR {each $ Clusters in Cluster $}                                
                   \STATE $ Clusters_{i} = member_{m_{i}} $
                   \STATE $ Distances_{i} = 0 $
                   \STATE $ Centroid_{i} = 0 $                    
       \ENDFOR

\COMMENT Iterations

\FOR {each members $ m_i$ in $G $}    
       \FOR {each members $ m_i$ in $Clusters $}
           \STATE $ Distances_{i} \leftarrow P_{member,member_{i}} $
               \IF {$ Distances_{i} \succ Centroid_{i} $}
                   \STATE $ Centroid_{i} \leftarrow Distances_{i} $
               \ENDIF
       \ENDFOR

\FOR {each $Centroid_{i}$}
       \IF {$ Centroid_{i} \succ MaxDistancesMax_{i} $}
       \STATE $ MaxDistancesMax_{i}  \leftarrow  Centroid_{i} $
       
       \IF {$ MaxDistancesMax_{i} \succ \eta $ AND $ ClustNB \neq -1 $}
   \STATE $Clusters_{i} = FUSION (Clusters_{i},Clusters_{NB})$
       \ENDIF
       \STATE $ ClustNB = i $
       \STATE $ Clusters_{ClustNB} \leftarrow member_{i}$
       \ENDIF    
\ENDFOR
\ENDFOR

\RETURN $ Clusters $
\end{algorithmic}
\end{algorithm}

\section{Conclusion and future work}

In this paper, we proposed a Friendlysourcing framework as a clustering based approach for data leak aware discovering competitive teams during the crowdsourcing  process in social network. First, the Markov model is used to estimate the hidden relationships between crowd members in the social network. Given the results of the previous step, then, the clustering approach groups the crowd members into data leak aware competitive teams.

In the future works, we plan to evaluate efficiency of the proposed approach by means of data leakage and time consumption.
Regarding the social network complexity, the current approach provides several classifications of competitive teams but not easy to choose the best one. Then, we will study how to take into consideration more constraints specifically user preferences.

\bibliographystyle{plain}
\bibliography{biblio,vldb_sample}
\end{document}